\documentclass[aps,twocolumn,superscriptaddress]{revtex4}
\usepackage{amsmath,amssymb}
\usepackage{graphics,graphicx}
\usepackage{dcolumn,bm}
\usepackage{psfrag}
\usepackage{color}
\usepackage{multirow,makecell}

\newcommand{\cu}
{\affiliation{Department of Physics, University of Calcutta,
92 Acharya Prafulla Chandra Road, Kolkata 700009, India.}}

\begin{document}

\title{
Non-equilibrium dynamics in Ising like models with biased initial condition}
\author{Reshmi Roy}
\cu
\author{Parongama Sen}%
\cu

\begin{abstract}
We investigate the dynamical fixed points of the zero temperature Glauber dynamics in Ising-like models.  
The stability analysis of the fixed points  in the mean field calculation 
shows the existence of an  exponent that depends on the coordination number $z$  in the Ising model.    
For the generalised voter model, a phase diagram is obtained based on this study.  
Numerical results for the Ising model for both the mean field case and short ranged models on lattices with different values of $z$ are also obtained. 
%; $x=0,0.5,1$ where $x=0.5$ is the unstable fixed point.
%The deviation $\delta$ from 0.5 follows the equation
%$\delta=\delta_0 \exp(\alpha t)$ where $\alpha$ depends on the 
%coordination number $z$. 
A related study is the  behaviour of the exit probability
  $E(x_0)$, defined as the probability that a configuration ends up with all spins up starting with $x_0$ fraction of up spins. An interesting result  is $E(x_0) = x_0$  in the  mean field approximation when $z=2$,  which is consistent with the conserved magnetisation in the system. For larger values of $z$,  $E(x_0)$ shows the usual  finite size dependent non linear behaviour both in the mean field model and  in Ising model  with nearest neighbour interaction on different two dimensional lattices. 
For such a behaviour, a  data collapse of $E(x_0)$ is obtained using $y = \frac{(x_0 - x_c)}{x_c}L^{1/\nu}$ as the scaling variable and $f(y)=\frac{1+\tanh(\lambda y)}{2}$ appears as the scaling function. The universality of the 
exponent and the scaling factor is investigated.

\end{abstract}

\maketitle

\section{Introduction}

Non-equilibrium dynamics  associated with  spin systems quenched from a
high temperature have 
been extensively studied in the past. Various  features associated 
with the ordering dynamics have been  explored for the  
 Ising model defined by the Hamiltonian 
$H= -J\sum _{ij} \sigma_i \sigma_j $ ($\sigma_i=\pm 1$) \cite{bray}. Classical spin models 
 have no intrinsic dynamics,
however, one can study the stochastic time evolution  using
certain  dynamical algorithms  that maintain the detailed balance \cite{krapivsky}. 
Glauber  dynamics is one of the popular choices
that reduces to a simple energy lowering scheme at zero temperature. 
% starting from a configuration far from equilibrium 
%\cite{bray}}. 
To study the ordering process, the system  is taken to be  completely disordered (i.e., at a high temperature) initially and suddenly cooled  to a lower temperature $T$; we  consider $T=0$ specifically in this paper.
In finite systems, the one dimensional Ising-Glauber  model, following  such a zero temperature quench,  
always ends up with all spins up or down
irrespective of the initial fraction of up spin $x_0$.
In higher dimensions, striped and blinkers state 
can also be reached  when the initial state is completely disordered, i.e., 
$x_0=0.5$ \cite{krapivsky,spirin1,spirin2,barros}.
On the other hand, there are a fairly large number of models which use Ising spins but without any
energy function associated with it, for example the Voter model. 
In such  models, the system  evolves by a given dynamical rule. 
%the dynamics where a spin follows the state of
%a randomly chosen neighbour. 

Various features  in the ordering process, 
for example domain growth, persistence, aging, time evolution of the order
parameter and other relevant quantities  have been   studied  for quite some time, 
particularly in the spin models. 
%In the ordering process, 
%One of the important features
  Exit probability is another  feature associated with the non-equilibrium dynamics  that has  received a fair amount of attention 
 more recently  \cite{stauffer,slan,lambi,crokidakis,claudio,pkm,sb_ps,pr_sb_ps,sb_ps_pr,prado,pr_ps15,timp,pm_ps16,pr_ps17,pm_ps17,sm_sb_ps}. The exit probability $E(x_0)$ is defined as the 
probability that a all-up configuration is reached starting from
$x_0$ fraction of up spins.   
$E(x_0)$ is linear in the one dimensional Ising model and the voter model (in all dimensions): $E(x_0)=x_0$
\cite{krapivsky}; this occurs due to the conservation of the order parameter. In contrast, 
 in the  two dimensional Ising system $E(x_0)$ is non-linear  and shows strong finite size effects \cite{pm_ps16}. The exit probability
as well as the dynamics have also been studied in the recent past for binary opinion dynamics models using 
mean field and several other analyses on hypergraphs, networks and simple square lattices \cite{slan,
lambi,prado,timp,noonan,moretti,galam}.

In this paper, we have considered the dynamics
of Ising and Ising like models where the evolution of the 
fraction of up spins ($x$) is studied. A mean field approach leads to the 
identification of the fixed points. We note that a non-trivial fixed point 
is $x= 0.5$ which corresponds to a disordered state. The stability 
of this fixed point is studied by starting from a biased
but uncorrelated initial condition where the initial fraction $x_0$ deviates 
from 0.5. The results for the mean field Ising model, obtained  for different values of $z$, the coordination number,  are compared with the short range model in finite dimensions. 
The fixed points for the generalised voter model (GVM) are also obtained parametrically and the 
 stability analysis   leads to obtaining the mean field  phase diagram in the two parameter plane.

%  Hence if $x(t)$ is the fraction of up spins
%at time $t$,  $x_0 \equiv x(0)  \neq 0.5$. 
%We have
%used mean field approach to calculate the fixed points 
%of the dynamics in the Ising model as well as in the generalised
% Voter model \cite{olivera_genvoter}. 

The evolution of $x$ in time helps in understanding  the behaviour of the 
exit probability. The exit probability is  computed numerically for the  Ising-Glauber model
in square, triangular lattices and the mean field Ising model. 
The results  for finite sizes show the existence of a scaling function with which two parameters can be associated, as noted in some earlier studies \cite{sb_ps,pr_sb_ps}.

\section {Mean field calculation in the Ising model}\subsection{Master equation approach} \label{sec-calc}

We first consider the Ising model in the mean field approximation. 
%  with different values of $z$, the coordination number. 
The master equation for the variable $x(t)$, the fraction of up spins at time $t$ is set up after calculating the spin flip probabilities.
%We consider a one dimensional lattice, where at each site, a spin variable ($\sigma= \pm 1$) is associated. 
%To study the dynamics of ordering in these systems, one starts with a initial fraction $x_0$ of up spin ($+1$) and 
%the rest of the spins are down having value $-1$. 
The system evolves under the zero temperature Glauber dynamics, i.e., 
spins are flipped when energy decreases by it and flipped with probability 1/2 
when energy does not change by flipping.  
For a particular spin, a neighbouring spin here is simply  another spin with which 
it interacts and  the number of such neighbours or the coordination number $z$ is taken as a variable. 
%ends
%up in achieving consensus state with all spins up or down depending on that 
%initial fraction of up spin $x$.

\subsubsection{z=2}
%When $z=2$,  we have effectively a one dimensional system. 

We first consider the case $z=2$.
Suppressing the argument $t$ for $x$,  an up spin flips with 
probability\\
(i) $(1-x)^2$ ,  when it has two neighbouring down spin\\
(ii)$2x(1-x)/2$ when it has 1 down neighbour and 1 up neighbour. This can happen in two ways and for each of the cases the spin flips with probability $\frac{1}{2}$.  

Denoting $P_+$ ($P_{-}$) as the total probability that a up (down) spin flips, one can therefore write, 

\begin{eqnarray}
P_+ & = &(1-x)^2 + x(1-x)\nonumber  \\
P_-&=&x^2 + x(1-x) 
\label{upflip2}
\end{eqnarray}
The evolution equation for $x(t)$ can be expressed in general as
\begin{eqnarray}
\frac{d x }{dt}&=&  (1-x) P_- -  x P_+,  
\label{mastereq}
\end{eqnarray}
which reduces to $\frac{d x }{dt}= 0$ using 
eq. (\ref{upflip2}).
This implies     
 $ x(t)  =x_0$, i.e., the dynamics conserve the 
magnetisation $m(t) = 2x(t) -1$  
%This is in consistency with the known result in one dimensional
%Ising Glauber model \cite{krapivsky}.
such that  $E(x_0) =x_0$ in this case obviously.

%{\color{blue} A finite Ising system with an initial fraction of $x_0$ up spin
%and a fraction of $1-x_0$ down spin leads to a initial magnetisation 
%$m_i=2x_0-1$. Finally, this system reaches consensus in which the system with
%magnetisation $m=1$ occurs with probability $E(x_0)$ and $m=-1$ occurs with probability $1-E(x_0)$. The final magnetisation is $m_f=E(x) \time 1 +(1-E(x)) \times (-1)=2E(x_0)-1$ which is same as initial magnetisation $m_i=2x_0-1$.
%Hence, the probability of achieving positive consensus is $E(x_0)=0$ and
%reaching negative consensus is $1-x_0$. }
%Therefore,  the conservation of magnetisation indicates that the exit probability is linear which does not depend on system size \cite{krapivsky}. 

%This average is over all trajectories of
%the dynamics.
%The exit probability $E(x)$ is defined
%as the probability that a configuration ends up with
%all spins up starting from a 
%initial fraction of up spin $x$ and for one dimensional Ising model $E(x_0)=x_0$.

%The square lattice Ising system evolving under
%zero temperature ends up in achieving consensus
% in $2/3$ of the total cases. In the remaining $1/3$ cases, frozen striped states are reached. 
%To study the dynamics of the Ising model in square lattice, an analytical calculation has been done

\subsubsection{z=4}

We next consider the case for $z=4$. 
In this case, an up spin flips with probability \\
(i) $(1-x)^4$ if it has 4 neighbouring down spins,\\  
(ii) $4x(1-x)^3$ when it has 3 down neighbours and 1 up neighbour which can happen in 4
ways, \\
(iii) $6 x^2(1-x)^2/2$ in case of 2 up and 2 down neighbours
(possible in 6 ways and the spin flips with probability $1/2$ in each case).\\ 
Therefore, 
\begin{eqnarray}
P_+  &=& (1-x)^4 + 4x(1-x)^3 + 3x^2(1-x)^2 \nonumber \\
P _- &=&  x^4 + 4x^3(1-x) + 3x^2(1-x)^2.
\label{upflip4}
\end{eqnarray}

%\begin{eqnarray}
%\frac{dx}{dt}= (1-x)P_- - x P_+,
%\label{mastereq}
%\end{eqnarray}

At the steady state, we  obtain from the master equation,
\begin{eqnarray}
\frac{dx}{dt}=-2x^3+3x^2-x=0, 
\label{steady_eq}
\end{eqnarray}
with the solutions   $x^*=0,0.5,1$.

To check the stability of the solutions, we  consider $x = x^*+\delta(t)$ where $\delta (t)$
is the deviation from the fixed point. For both $x^*=0$ and $1$,
considering only up to  linear order terms in $\delta$,
we get, 
$\frac{d\delta}{dt}=  - \delta$,
the solution of which is 
\begin{equation}
\delta (t) = \delta_0 \exp(-t).
\end{equation}
where $\delta_0 \equiv \delta(t=0)$.
The negative exponent 
implies that $x^*=0$ and 1
are stable fixed points. 
%{\color{blue} This exponent can be compared to the
%Lyapunov exponent. However, the Lyapunov exponent is related to the rate of growth of the distance between trajectories which start very close in system undergoing deterministic evolution and the exponent involves a large time limit. In the present work, the time evolution of a stochastic system is discussed for which the same initial condition would lead different trajectories whose distance growth with time and the exponent characterises the rate of growth of up spin for a short time range. Hence, the exponent is not exactly Lyapunov exponent, rather a Lyapunov like short time exponent.}   
 
For $x^*=0.5$, one gets 
\begin{eqnarray}
\delta(t) =\delta_0\exp{\bigg(\frac{1}{2}t\bigg)}. 
\end{eqnarray}
The positive exponent here implies that $x^*=0.5$ is an unstable fixed point. 
Of course, $\delta$ cannot increase indefinitely and its extreme  values are 
$\pm 0.5$.
The stability analysis thus 
shows that  the system ends up with all spins up/down  (for $\delta_0$ positive/negative). 
The magnetisation $m(t) = 2\delta(t)$ here. 

\begin{figure}[h]
\includegraphics[width=8cm]{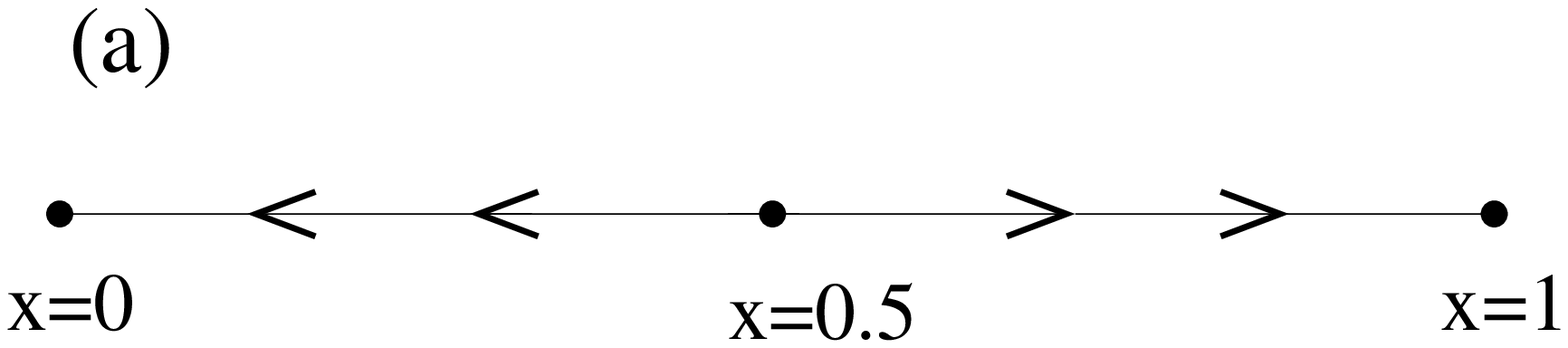}
\includegraphics[width=8cm]{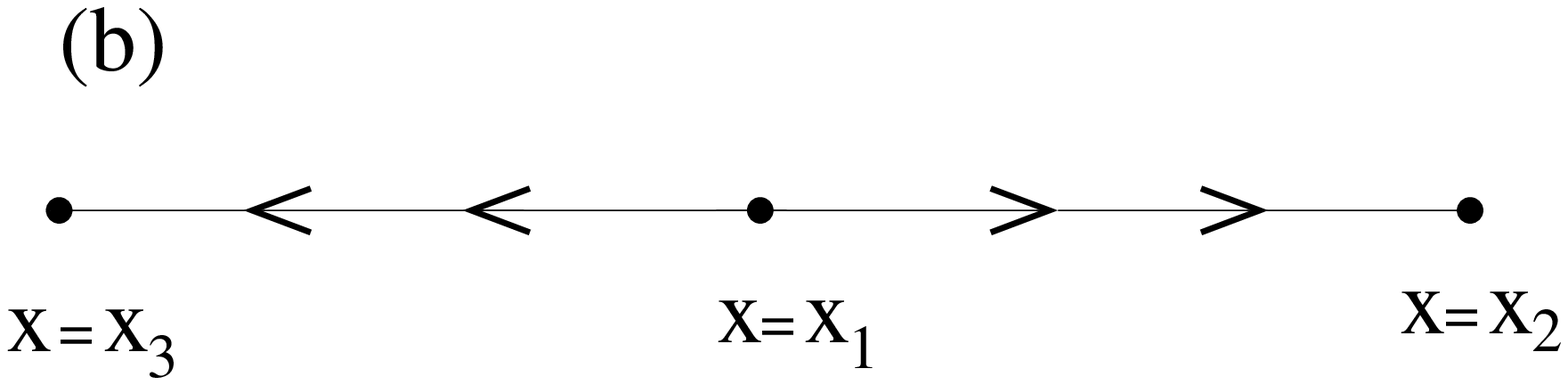}
\caption{(a) Flow diagram for the mean field Ising model for $z > 2$. 
$x=0$ and 1 are the stable fixed points and $x=0.5$ is unstable. (b) shows
the flow diagram for the generalised voter model. $x_1=0.5$ is the
unstable fixed point and $x_2=\frac{1}{2}+\frac{1}{2} \sqrt{\frac{2z_1+z_2-2}{2z_1-z_2}}$
and $x_3=\frac{1}{2}-\frac{1}{2} \sqrt{\frac{2z_1+z_2-2}{2z_1-z_2}}$ are stable fixed points
provided that the  exponent is positive.
}
\label{stability}
\end{figure}

\subsubsection{z=6}

A similar analysis is done for $z=6$. Here, $P_+$ and $P_-$ can be expressed as
\begin{eqnarray}
P_+ & =& (1-x)^6 + 6x(1-x)^5 + 15x^2(1-x)^4 + 10x^3(1-x)^3 \nonumber\\
P_-& = &x^6 + 6x^5(1-x) + 15x^4(1-x)^2+ 10x^3(1-x)^3 \nonumber \\
%+ 10x^3(1-x)^3
\label{upflip3d}
\end{eqnarray}
and 
therefore the master equation can be written as 
\begin{eqnarray}
\frac{dx}{dt}=  6x^5-15x^4+10x^3-x.
\label{master3d_eq}
\end{eqnarray}
Solving the steady state equation $\frac{dx}{dt}$=0,
one gets, $x^*=0,\frac{1}{2},\frac{3+\sqrt 21}{6},\frac{3-\sqrt 21}{6}, 1$.
The third and fourth solution being unphysical, $x^*=0,0.5$ and $1$ are the only physical solutions.
Considering $x = x^*+\delta (t)$ in  
Eq. (\ref{master3d_eq}), the solution becomes $\delta (t)=\delta_0 \exp(-t)$ for $x^*=0$ and 1
which are stable fixed points. For $x^*=0.5$, one obtains
\begin{eqnarray}
\delta(t) = \delta_0\exp\bigg(\frac{7}{8}t\bigg)
\label{3d_delta}
\end{eqnarray}
which  shows that $x^*=0.5$
is again an unstable fixed point. 
 The larger value of the exponent in the $z=6$ case indicates that 
%if $x \to 0.5^+$, the positive consensus is achieved much faster than the $$ case.
the dynamics are faster for $z=6$ compared to that in the  $z=4$ case. Fig. \ref{stability}a
shows the flow diagram for the mean field Ising model for $z=4$ and $z=6$.

From the above studies, we conclude that in  the mean field approximation, in general $\delta(t) = \delta_0 \exp ({\gamma t})$ where $\gamma$ increases with $z$. 
This behaviour is a short time one as the system reaches the stable fixed points  at long times,   
confirmed by the simulation results discussed in the next subsection.  Hence  
$\gamma $ can be interpreted as an  inverse time scale over which the exponential growth 
of  $\delta(t)$ can be observed.

\subsection {Simulation results} \label{sim-results}

To check the mean field results we have conducted simulations where 
a spin interacts with randomly chosen $z$ neighbours. The  system consists of $N$ spins and the  choice of the random neighbour is made in an annealed manner which implies the 
interaction can take place with different spins at each step in general. 

%Although the geometry is not important, we have  
%studied the dynamics on  $L\times L$ lattices in general. 
We defer the discussion on the $z=2$ case  to   section \ref{exitsection} and 
consider the cases $z=4$ and 6  where  
we expect an unstable point at $x = 0.5$.
We have started from a fixed initial fraction of up spin $x_0 = 0.5+\delta_0$ 
with $\delta_0 > 0$ and studied
how the fraction  $\delta (t)$ $(=x(t)-0.5)$, evolves in time. $N$ updates constitute one
single Monte Carlo step. Here, we have considered only those configurations
for which positive consensus is attained
to obtain  the  exponent $\gamma$ and compare with the result found in  the analytical calculation. 

\begin{figure}[h]
\includegraphics[width=8cm]{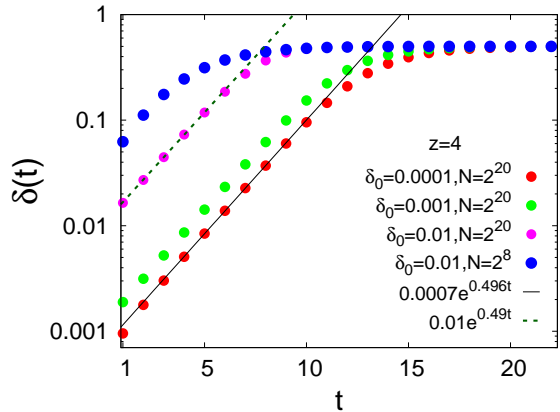}
\caption{Variation of $\delta(t)$ are shown with time  for $z=4$ for different $\delta_0$ where mean field approach is used. This plot also shows the data for
several system sizes. 
Data are fitted to the exponential function, mentioned in the key. Maximum number of configuration was 5000. $\delta(t)$
attains the saturation value faster for larger $\delta_0$ 
and the process is slower 
for smaller value of $\delta_0$.}
\label{mean4}
\end{figure}

\begin{figure}[h]
\includegraphics[width=7.5cm]{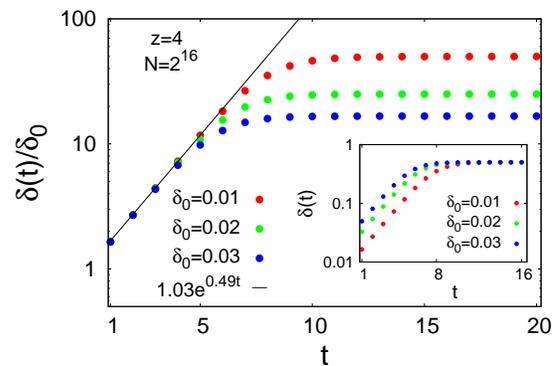}
\caption{Data collapse of $\delta(t)$ are shown with time for several $\delta_0$ where $z=4$ and data are fitted to
an exponential form as mentioned in the key. 
These data are for a system of $2^{16}$ spins averaged over a maximum of 5000 realisations. Inset shows the 
unscaled data. In this simulation mean field approach is used.}
\label{mean4_scale}
\end{figure}

$\delta (t) $ shows an exponential growth with time which  shows consistency with 
the results of  section \ref{sec-calc} as $N$ is increased. The results for $z=4 $  shown 
in Fig. \ref{mean4} indicate the exponential growth becomes more noticeable as $N$ increases and that the associated exponent $\approx 0.5$ is  independent of  
 $\delta_0$ for all practical purposes. 
A data collapse for different values of $\delta_0$ is obtained by scaling $\delta(t)$
by $\delta_0$, shown in Fig. \ref{mean4_scale} which is also consistent with the analytical results.

\begin{figure}[h]
\includegraphics[width=8.5cm]{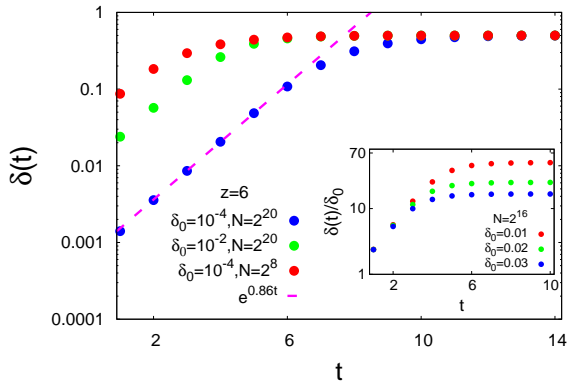}
\caption{Plots of $\delta(t)$ are shown against time for several system sizes where $z=6$. Data for several values of $\delta_0$
are also shown. 
Inset shows the data collapse for different $\delta_0$ for a system of $2^{16}$ spins. As $\delta_0$ increases, $\delta(t)$ rapidly
saturates. Minimum Number
of different initial configuration was 2000 and mean field approximation is used
for this simulation.}
\label{mean6}
\end{figure}

A similar estimation has been done for $z=6$ by considering the interaction of the selected spin
with  randomly chosen 6 neighbours. $\delta(t)$ shows an exponential behaviour with time
again and 
 the exponent is $\sim 0.86$. 
Fig. \ref{mean6} shows the results.
The exponent $7/8$
obtained in section \ref{sec-calc} for $z=6$  agrees fairly well with the simulation results.  

It should be mentioned here that  the saturation is obtained very rapidly, the exponential fitting is therefore valid only for a few initial 
time steps. The saturation is enhanced for larger values of $\delta_0$ and $z$.

\section{ Simulations for short range  models on lattices}

The  simulations for the Ising model are  repeated  on  two dimensional  lattices  where the spins have short range interactions. 
 We consider the vicinity of the unstable fixed point again, such that $x(0) = 0.5 + \delta (0)$ and study the evolution of $\delta(t)$ where $x(t) = 0.5 + \delta(t)$.  

In order to check the dependence on $z$, we have considered square lattices with nearest and nearest plus next nearest neighbours and triangular lattices with nearest neighbours such that   $z=4, 8 $ and 6 respectively.

It is well known that the absolute value of the magnetisation grows as $t^\beta$ in the ordering process of the Ising model in all finite dimensions when the initial configuration is completely disordered. 
This follows from the fact that domains of up/down spins both grow as $t^{\eta d}$ where $\eta$ is the domain growth exponent and d is the spatial dimension. 
The magnetisation is given by  $m = \sum \xi_i$, where $\xi_i \propto \pm t^{\eta d}$ are uncorrelated random variables and the sum is over all domains. The stochastic variable $m$ thus 
satisfies $\langle m \rangle =0$ and $\langle m^2\rangle \propto t^{\eta  d}$, leading to the   result  $|m| \propto  t^{\frac{\eta d}{2}}$. One can also derive this  from the dynamic scaling obeyed by the correlation function \cite{bray}. 
It is known that   
$\eta=\frac{1}{2}$ in all dimensions and thus $\beta$ is dependent on the dimension;  in two dimensions  $\beta = 0.5$. 
For $x(t) = 0.5 +\delta(t)$, as mentioned before, magnetisation is simply $2\delta (t)$ and the variation  of $m(t)$ and $\delta(t)$ would be identical.  
%The magnetisation is thus identical to $2 \delta(t)$.
%Here, the

%contribution from 4 nearest neighbour is taken into account. A spin flips depending on the 

%energy minimization scheme according to the zero temperature Ising Glauber dynamics. 
%When a spin flip costs energy $\Delta E$,
%(i) if $\Delta E>0$ that spin will flip   
%(ii) if $\Delta E=0$  the spin flips with probability $1/2$  
%(iii) if $\Delta E>0$, the spin does not flip. 

%\subsection {z=4}

\begin{figure}[h]
\includegraphics[width=8cm]{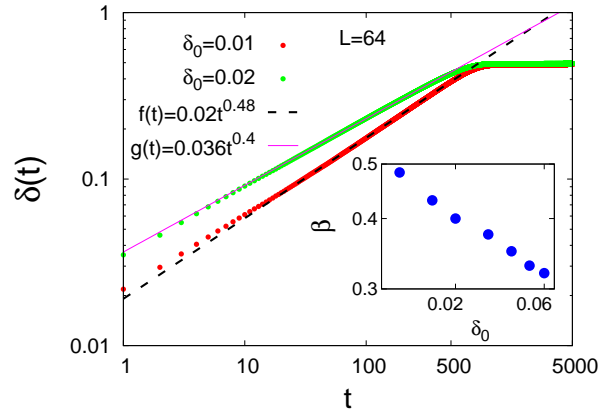}
\caption{Variation of $\delta(t)$ with time is shown in square lattice nearest neighbour Ising model
for several $\delta_0$. Data are fitted to the power law
form, exponents are mentioned in the key. The power law exponent $\beta$ decreases as the value
of $\delta_0$ increases. Inset shows the variation of $\beta$ with $\delta_0$.
These data are for system size $L \times L=64 \times 64$
averaged over 5000 realisations.}
\label{ising2d}
\end{figure}

It is observed that for any value of $z$ and $\delta_0$,  $\delta(t)$ shows a power law behaviour with time before reaching the saturation value for all values of $\delta_0$;

\begin{eqnarray}
\delta(t) \sim  t^{\beta}.
\label{ising_eq}
\end{eqnarray} 
The results for $z=4$ are shown in Fig. \ref{ising2d}. 
 The value of $\beta$ depends on $\delta_0$, 
 as $\delta_0$ increases (which means  the system is more ordered to begin with),   
it decreases as 
shown in the inset of Fig. \ref{ising2d} for $z=4$. This is understandable, in the limit  $\delta_0 \to 0.5$, the system  is almost static such that the time dependence is weak reflected by a smaller value of $\beta$.

\begin{figure}[h]
\includegraphics[width=8cm]{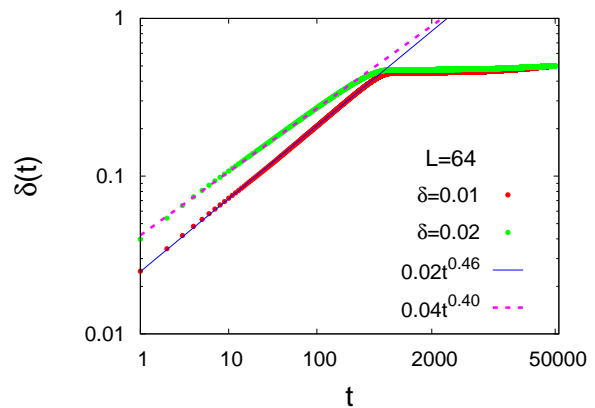}
\caption{Variation of $\delta(t)$ with time is shown in triangular lattice
for nearest neighbour interaction. Data are fitted to the Eq. \ref{ising_eq}
and the  exponents are mentioned in the key.
These data are for system size $L \times L=64 \times 64$
averaged over 5000 realisations.}
\label{ising_tri}
\end{figure}

In the triangular lattice, where $z=6$, $\delta(t)$ is also found to show a
power law variation with time according to Eq. \ref{ising_eq}.
 As $\delta_0$ increases, $\beta$ decreases as indicated by the data presented in   Fig. \ref{ising_tri}.
The values of $\beta$ are reasonably close to those obtained in the square lattice. 

\begin{figure}[h]
\includegraphics[width=7cm]{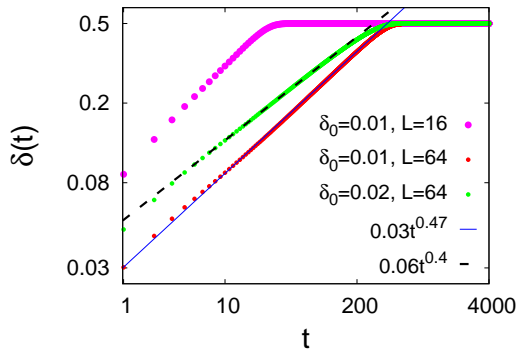}
\caption{Variation of $\delta(t)$ with time is shown in square lattice next nearest neighbour 
Ising model for several system sizes. Data are fitted to the
power law form as mentioned in the key.}
\label{nn_ising}
\end{figure}

We  also consider the  the Ising model with a Moore neighbourhood where
next nearest neighbour interactions are included and $z=8$. 
The Hamiltonian of this system 
is given by
\begin{eqnarray}
H = -J_1 \sum \limits_{<i,j>} S_iS_j -J_2 \sum \limits_{<i,j'>} S_iS_j,
\label{ising_nneq}
\end{eqnarray}
where $J_1$ and $J_2$ are the strengths of interaction for nearest
neighbour and next nearest neighbour respectively.
We have considered  the interactions to be equal in strength, $J_1 = J_2$.
Here, $z=8$ and 
once again we find a behaviour similar to $z=4,6$  in two dimensions
(see Fig. \ref{nn_ising}).

\begin{figure}[h]
\includegraphics[width=7cm]{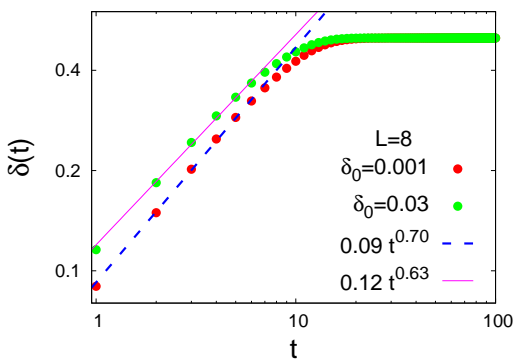}
\caption{Variation of $\delta(t)$ with time is shown in cubic lattice nearest neighbour Ising model
for different $\delta_0$. $\beta$ decreases as the value
of $\delta_0$ increases. Data are fitted to the power law
form, exponents are mentioned in the key. The power law exponent $\beta$ decreases as the value
of $\delta_0$ increases.
These data are for system size $L^3=8^3$
averaged over 5000 realisations.}
\label{ising3d}
\end{figure}

It is also interesting to check whether for the same value of $z$ but in a different dimension, the value of $\beta$ remains the same. For this, 
simulations have  been conducted on cubic lattice Ising
system where $z=6$ as in the triangular lattice.  $\delta(t)$ shows a power law variation in this case also; however, 
the exponent $\beta$ is larger compared to the two dimensional case 
(see Fig. \ref{ising3d}).

The above results show that the exponent $\beta$ is independent of $z$ 
in two dimensions while for three dimensions, with the same $z$ we find a different value of $\beta$ when $\delta_0 \neq 0$.

\section{Exit probability}

In this section, we present the results for $E(x_0)$, the probability that the system ends up in a state with all spins up,    starting from an initial state with $x_0$ fraction of up spins. 
Since some results are already known for the short range Ising models, we first 
discuss that and then continue to report the results for the mean field case.

\label{exitsection}

\subsection{Results for nearest neighbour interactions}
Next, we have studied the exit probability for the two dimensional 
nearest neighbour Ising models.
Exit probability $E(x_0)$ is known to have a liner behaviour $E(x_0)=x_0$ for 
one dimensional Ising Glauber model. In two dimensional model $E(x_0)$ is non linear
and shows  strong finite size effects \cite{pm_ps16,pm_ps17,pr_ps17}. As the system size increases the curves become steeper and 
the data suggest that $E(x_0)$ approaches a step function
in the thermodynamic limit. Finite size scaling can be done
using the form 
\begin{eqnarray}
E(x_0,L)=f[\frac{x_0-x_c}{x_c}L^{\frac{1}{\nu}}]
\end{eqnarray}
as observed  in \cite{sb_ps}, where $f(y) \to 0$ for $y \ll 0$ and is equal to 1 for $y \gg 0$. 
Therefore, a data collapse for different system size can be obtained when $E(x_0)$ is 
plotted against $\frac{x_0-x_c}{x_c}L^{1/\nu}$ where $x_c=0.5$. 
%Fig. \ref{exit_probsquare} shows the best data collapse for a square lattice Ising model 
%when $\nu=1.3$
On square lattices, an Ising system freezes into a striped configuration for $x_0=0.5$ in  $33.9$ percent cases (an exact result \cite{barros}) in the thermodynamic limit. Numerical simulations show that   the   freezing probability  has   strong system size dependence \cite{spirin1,spirin2}. However, 
the dynamical scaling behaviour remains intact in spite of the freezing. 
Very close to  $x_0=0.5$, such
frozen striped states may occur with a nonzero probability 
in finite systems as shown in 
\cite{spirin2}. 
% reached in some configurations which is a considerable fraction (lower than 33.9\%) for finite system  
While calculating $E(x_0)$, 
such configurations have been discarded. 

The data collapse is obtained using eye estimation 
for square lattice Ising model when $\nu \approx 1.3$ agreeing with the result of \cite{pr_ps17,pm_ps17}. 
The collapsed data can be fitted to the form
\begin{eqnarray}
f(y) = \frac{1 + \tanh(\lambda y)}{2}
\label{scale_hfunc}
\end{eqnarray}
as in \cite{pr_sb_ps}. The value of $\lambda$ turns out to be 1.10 using GNUFIT.

To get a more accurate
value of $\nu$  required for obtaining best data collapse, we have employed another method  
used previously in \cite{pm_ps17}. We have calculated $y=\frac{x_0-x_c}{x_c}L^{1/\nu}$
for the different values of $L$. As the data collapse are supposed to fit to the form of Eq. \ref{scale_hfunc}
we have chosen the range of $\nu$ and $\lambda$ for which the collapse and fitting seem good.
We have varied the values of $\nu$ and $\lambda$ in steps of 0.001 and for every pair we have
calculate the error $\epsilon$ given by

\begin{eqnarray}
\epsilon=\frac{1}{n}\sqrt{\sum\limits_n {[f(y)-E(x_0)]}^2}
\label{error}
\end{eqnarray}

\begin{figure}[h]
\includegraphics[width=8cm]{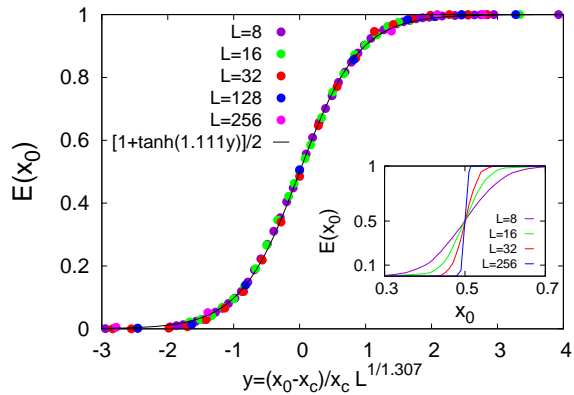}
\caption{Data collapse of $E(x_0)$ is shown for different 
system sizes in square lattice Ising model using $\nu=1.307$. Data are fitted to the form
of Eq. \ref{scale_hfunc} as mentioned in the key.
 Inset shows the unscaled data. Number
of different initial configuration was 5000 for these simulations. }
\label{exit_probsquare}
\end{figure}

The pair of $\nu$ and $\lambda$ for which the minimum value of $\epsilon$
is obtained are the optimal values required for best data collapse and scaling function.
The values of $\nu$ and $\lambda$ are 1.307 and 1.111 using this method and the results are shown in
Fig. \ref{exit_probsquare}.

\begin{figure}[h]
\includegraphics[width=8cm]{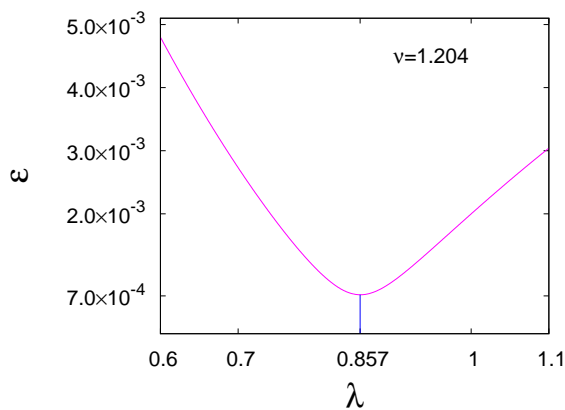}
\caption{Variation of the least square error $\epsilon$ with $\lambda$ is shown
for $\nu=1.204$ in triangular lattice. The minima of the curve is at $\lambda=0.857$.}
\label{error_plot}
\end{figure}

\begin{figure}[h]
\includegraphics[width=7cm]{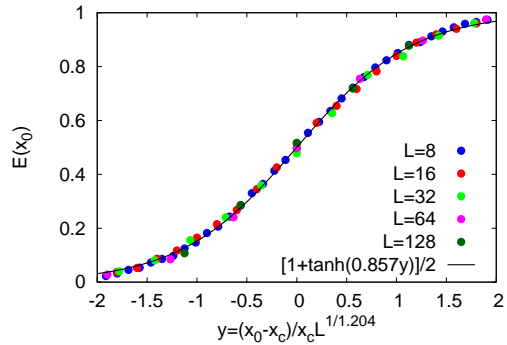}
\caption{Data collapse of $E(x_0)$ is shown for different system sizes in triangular lattice using 
$\nu=1.204$ and data 
are fitted to Eq. (\ref{scale_hfunc}).  Number
of different initial configuration was 5000 for these simulations.}
\label{exit_probtri}
\end{figure}

These results were already available from previous studies,  although for smaller system sizes.  We repeat these simulations as  our aim is to 
 determine if any universality in the scaling behaviour exists in two dimensional Ising
systems. Hence we have studied the exit probability in a triangular lattice (number of nearest neighbours $z=6)$. To obtain the best data collapse the least square method
has been employed in this case also and   is
graphically illustrated in Fig. \ref{error_plot}. The data collapse of $E(x_0)$ for different 
system sizes is obtained with $\nu=1.204$ using the above method and the scaled data
is fitted according to Eq. (\ref{scale_hfunc}). Fig. \ref{exit_probtri} shows the results.
The value of 
$\nu$ is close but not exactly  equal to the 
value obtained for square lattice. $\lambda = 0.857$ is definitely different.  

%$E(x_0)$ shows a similar behaviour in cubic lattice Ising
%model where a spin can interact with its 6 nearest neighbours. However,
%the curves are steeper when compared to square lattice of same system size
%and the data collapse for different system sizes indicates that $\nu$ is also different
%from the two dimensional case; $\nu \approx 0.47$.

%\begin{figure}[h]
%\includegraphics[width=7cm]{3d_scale.eps}
%\caption{Data collapse of $E(x_0)$ is shown for cubic lattice Ising model for different system sizes. Number
%of different initial configuration was 2000 for these simulations. The data are fitted to Eq. (\ref{scale_hfunc}) 
%with $\lambda=0.4$ as
%mentioned in the key.}
%\label{exit_cube}
%\end{figure}

\subsection{Results for mean field like model}

Here we present the results for the exit probability 
$E(x_0)$  using mean field approach 
where the $z$ neighbours are chosen randomly. 

\begin{figure}[h]
\includegraphics[width=7cm]{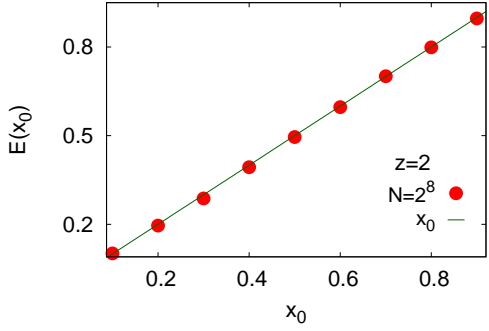}
\caption{$E(x_0)$ is shown against $x_0$ for $z=2$ where mean field approach
is used. $E(x_0)$ shows a linear variation with $x_0$. Number
of different initial configuration was 5000 for these simulations.}
\label{exit_mean2}
\end{figure}

For $z=2$,  the exit probability  shows a linear behaviour
$E(x_0) = x_0$ (see Fig. \ref{exit_mean2}). This is in consistency with the conservation we noted for $x$ in section \ref{sec-calc}. 
%For comparison, we also show the exit probability for the 
%one dimensional Ising model where the linearity occurs due to conservation. 
It may seem  surprising that the mean field result  with $z=2$ gives the 
exact result known for the one dimensional Ising model.  We attempt to justify why this happens in the following way.

We note that for the voter model, the $i$th spin $\sigma_i$ flips with a probability \begin{equation}
w(\sigma_i) = \frac{1}{2}[1-\sigma_i\sum_j\sigma_j/z]
\end{equation}
where $j$ is a neighbour of $i$. This probability is  valid in any dimension. 
Thus the above dynamics in the voter model conserve the 
total spin in any dimension. 
It is well known that in one dimension, 
the voter model  dynamics  coincide with the Ising dynamics where $z=2$.
In the  
 mean field calculations for $z=2$ it is evident that  the voter model dynamics are  being used precisely and
since in the latter, conservation is valid always, we get a result which is the exact one for the one dimensional Ising model too. 
It is interesting to note that hence for the $z=2$ case, it does not matter whether one picks up randomly any two neighbours  or strictly the nearest neighbours as far as conservation is concerned. 
We have also checked that if the choice of neighbors is done randomly in a quenched manner, the results remain the same.

%To get a variation of fraction of up spin $\delta$ for $z=2$ case,
%we have started from a initial fraction of up spin $\delta (=x-0.5)$ as done
%in section \ref{sim-results} which shows that at early time there is a
%fluctuation about $\delta_0$, then $\delta$ saturates to $\delta_0$. Here, to calculate
%$\delta$, all the configurations are considered which ends up with all spins up or down.

%\begin{figure}[h]
%\includegraphics[width=8cm]{delta_z2.eps}
%\caption{Variation of $\delta$ are shown with time  for $z=2$ for $\delta_0=0.01$. 
%These data are averaged over 5000 different realisations.}
%\label{mean2}
%\end{figure}

\begin{figure}[h]
\includegraphics[width=8.5cm]{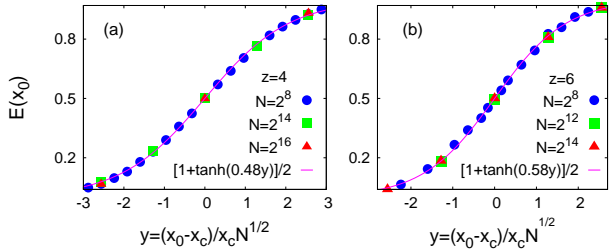}
\caption{(a) Data collapse of $E(x_0)$ is shown for different system sizes using $\nu^\prime=2$ where $z=4$. (b) shows the data collapse of $E(x_0)$ for different system sizes with $\nu^\prime=2$ for $z=6$. Mean field approximation is used in these simulations where number of different initial configuration was 5000. Data are fitted to the functions as mentioned in the key.}
\label{exit_mean4}
\end{figure}

For other values of $z$, $E(x_0)$ becomes steeper in the mean
field case than that was obtained using nearest neighbour interactions. Here, a data collapse is obtained 
by plotting $E(x_0)$ against $\frac{x_0-x_c}{x_c}N^{1/\nu^\prime}$
(where  $N$ is the total number of spins)
using $\nu^\prime =2$  for both $z=4$ and $z=6$. 
The scaled data are fitted to the form of
Eq. (\ref{scale_hfunc}) where $y=\frac{x_0-x_c}{x_c}N^{1/\nu^\prime}$.  
The value of $\lambda$ obtained for $z=4$ is $\approx 0.48$ and
$\lambda \approx 0.58$ when $z=6$. Data collapse of $E(x_0)$ is shown 
in Fig. \ref{exit_mean4} for $z=4$ and $z=6$. 

Table \ref{table} shows
the values of $\nu$, $\nu^\prime$ and $\lambda$ obtained numerically
for the  
short ranged and mean field 
 Ising  models. Note that for the short ranged systems, $\nu^\prime = 2 \nu$. 
These results are discussed in the last section. 

\begin {table}[h]
\caption {$\nu$, $\nu^\prime$ and $\lambda$ obtained for Ising model using numerical simulations.}
\begin{center}
\begin{tabular}{ |c|c|c|c|c| }
\hline
\multirow{3}{*}{Quantity} & \multicolumn{2}{c|} {\thead{Nearest neighbour (NN) \\ interaction}} &  \multicolumn{2}{c|} {Mean field} \\ 			

\cline{2-5}
	&	 	&		&		&		\\
  	&  Square 	& Triangular  	& 	z=4	& 	z=6 	\\
  	&		&		&		&		\\
\hline
              &	 	&		&		&		\\
 $\nu,\nu^\prime$ & $\nu=1.307(1)$		& 	$\nu=1.204(1)$	&  	$\nu^\prime\sim 2$	&  $\nu^\prime\sim 2$ 	\\
 ($\nu^\prime=2\nu$ for         &	 	&		&	 	&	   			\\
  NN models)        &	 	&		&		&					\\
       &	 	&		&			&						\\
\hline
              &	 	&		&		&		\\
$\lambda$ & 1.111(1)	& 	0.857(1)	 & 	$\sim 0.48$	& $\sim 0.58 $  	\\
	&	 	&		&		&		\\
\hline

\hline

\end{tabular}
\label{table}
\end{center}
\end {table}
%\end{small}

\section{Generalised voter model (GVM)}

%To get a deeper insight about the dynamics, 
We have considered next the generalised voter model.
We first describe the model on a square lattice where 
a spin variable $\sigma_i = \pm 1$ is associated 
with every site of the lattice. The time
evolution is governed by a  
single spin flip stochastic dynamics;
the spin flip probability $w_i(\sigma)$ for
the $i$-th spin is given by \cite{olivera_genvoter},

\begin{eqnarray}
w_i(\sigma) = \frac{1}{2}[1 - \sigma_if_i(\sigma)],
\end{eqnarray}
where $f_i(\sigma) = f(\Sigma_\delta \hspace{1mm}\sigma_{i+\delta})$,
a function of the sum of
the nearest neighbor spin variables. The model is defined
taking $f(0) = 0$, $f(2) = -f(-2) = z_1$ and $f(4) =
-f(-4) = z_2$, where $z_1$ and $z_2$ are restricted to  
$z_1 \leq 1$ and $z_2 \leq 1$. The original voter model
corresponds to $z_1=0.5$ and $z_2=1$ and Ising model
is recovered for $z_1=1,z_2=1$.

In the  mean field approximation  to obtain the master equation, 
the above  dynamical rule is followed which means that $z=4$ is taken and the parameters defined as above. 
%an up spin flips in the following cases when it has:\\
%(i) 4 up neighbouring spin; corresponding probability  $P_1= \frac{1}{2}(1-z_2)x^4$,\\
%(ii) 3 up neighbouring spins and one down neighbour $P_2= 2(1-z_1)x^3(1-x)$,\\
%(iii) 2 up neighbours and 2 down neighbour a$P_3= 3x^2(1-x)^2$,\\
%(iv) 1 up neighbour and 3 down neighbours $P_4= 2(1+z_1)x(1-x)^3$,\\
%(v) 4 down neighbours $P_5=\frac{1}{2}(1+z_2)(1-x)^4$. \\
For an up spin,  flipping  probabilities  $P_i ~(i= 0~ {\rm to}~ 4)$, when there are $i$  neighbouring spins 
in the up state are given by 
\begin{eqnarray}
P_1&=& \frac{1}{2}(1-z_2)x^4, \nonumber\\
 P_2&=& 2(1-z_1)x^3(1-x), \nonumber\\
 P_3&=& 3x^2(1-x)^2, \nonumber\\
 P_4&= &2(1+z_1)x(1-x)^3,\nonumber\\
P_0&=&\frac{1}{2}(1+z_2)(1-x)^4. \nonumber
\end{eqnarray}

The total probability $P_+$ that an up spin flips is   
$P_+ = \sum_{i=0}^{4}P_i$  such that    
\begin{eqnarray} %P_+ = P_1+P_2+P_3+P_4+P_5 \nonumber\\ 
P_+= \frac{1}{2}(1-z_2)x^4+2(1-z_1)x^3(1-x) \nonumber\\
+ 3x^2(1-x)^2 + 2(1+z_1)x(1-x)^3 \nonumber\\
+ \frac{1}{2}(1+z_2)(1-x)^4 . 
\label{upflip}
\end{eqnarray}
Similarly  the probability $P_-$ that a down spin flips is 
\begin{eqnarray}
P_- =  \frac{1}{2}(1-z_2)(1-x)^4+2(1-z_1)x(1-x)^3 \nonumber\\
+ 3x^2(1-x)^2 + 2(1+z_1)x^3(1-x) + \frac{1}{2}(1+z_2)x^4. 
\label{downflip}
\end{eqnarray}

%Therefore, the master equation is
%\begin{eqnarray}
%\frac{dx}{dt} = (1-x)P_--xP_+ = 0 \\
%\implies \frac{dx}{dt} = x^3(-4z_1+2z_2)+x^2(6z_1-3z_2) \nonumber\\
%+x(-2z_1+2z_2-1)+\frac{1}{2}-\frac{z_2}{2}=0
%\label{voter_eq}
%\end{eqnarray}

Therefore, the master equation $\frac{dx}{dt} = (1-x)P_- - xP_+$
reduces to
\begin{eqnarray}
 \frac{dx}{dt} = x^3(-4z_1+2z_2)+x^2(6z_1-3z_2) \nonumber\\
+x(-2z_1+2z_2-1)+\frac{1}{2}-\frac{z_2}{2}.
\label{voter_eq}
\end{eqnarray}

Putting the values $z_1=0.5$ and $z_2=1$ in eq. \ref{voter_eq}, one gets 
%\begin{eqnarray}
$\frac{dx}{dt}= 0$,   
 consistent with the voter model result that there is conservation in any dimension. 
On the other hand, 
by taking  $z_1=1$ and $z_2=1$ in Eq. \ref{voter_eq}, 
Eq. \ref{steady_eq} is recovered for the Ising model with $z=4$.

For general values of $z_1$ and $z_2$, the steady state condition leads to three fixed points $x_1, x_2, x_3$ where 
$x_1=0.5$, $x_2=\frac{1}{2}+\frac{1}{2} \sqrt{\frac{2z_1+z_2-2}{2z_1-z_2}}$  and  $x_3=\frac{1}{2}-\frac{1}{2} \sqrt{\frac{2z_1+z_2-2}{2z_1-z_2}}$.
Now, let us take $x(t) =  x_1 + \delta(t)$, i.e., the behaviour close to the fixed point $x_1 = 0.5$.
%\begin{eqnarray}
%\frac{d(x+\delta)}{dt} = (x+\delta)^3(-4z_1+2z_2)+(x+\delta)^2(6z_1-3z_2) \nonumber\\
%+(x+\delta)(-2z_1+2z_2-1)+\frac{1}{2}-\frac{z_2}{2}=0 \nonumber.
%\end{eqnarray}
%\implies \frac{d\delta}{dt} = \delta(z_1+\frac{1}{2}z_2-1) \nonumber \\
Considering up to   linear term in $\delta$ only, $\delta(t)$ is found to be 
\begin{eqnarray}
\delta (t)  = \delta_0 \exp (z_1+\frac{1}{2}z_2-1)t, 
\label{voter_stability}
\end{eqnarray}
where $\delta_0 \equiv \delta (t=0)$.
%\delta &=&\delta_0 \hspace{0.5cm} \rm{for\hspace{0.2cm}original\hspace{0.2cm} voter \hspace{0.2cm} model}
The exponent is thus  $z_1+\frac{1}{2}z_2-1$.

%\begin{figure}[h]
%\includegraphics[width=10cm]{voter.eps}
%\cation{}
%\label{voter}
%\end{figure}

We will now analyse the  sign of the exponent and thus the stability of the fixed point $x_1 = 0.5$ which 
corresponds to a completely disordered state. Since the magnetisation $m$ is given by $2x -1$, 
%Magnetisation $ m_0= 2x-1 = 0$ for $x=\frac{1}{2}$.
$|m| = (\frac{2z_1+z_2-2}{2z_1-z_2})^{\frac{1}{2}}$ for $x_2$ and $x_3$. 
$m$ can have non-zero values for $x_2$ and $x_3$, provided $ \frac{2z_1+z_2-2}{2z_1-z_2} \geq 0$ and also we
require 
$|m| \leq 1$.
The first criterion is satisfied ($|m| > 0$)   when either  (i) $2z_1+z_2-2 \geq 0$
and $2z_1-z_2 > 0$ or when (ii) $2z_1+z_2-2 \leq 0$
and $2z_1-z_2 < 0$. We note in the first case the first condition implies the second and in the next case the second condition implies the first one. 
Hence, for $|m| >  0$,  
$2z_1 +z_2 -2$ and $2z_1-z_2$ can in principle be either both positive or both negative. However,
the condition that $|m| \leq 1$ is violated for  case (ii) since $z_1, z_2 \leq 1$ and hence $m =0$ is the only possible solution when  $2z_1 + z_2 -2  < 0$. 
Thus the only condition for an ordered region to exist is that the 
quantity $2z_1 + z_2 -2$ must be positive.   This is consistent with the fact that
the exponent (which is an identical expression in $z_1, z_2$), has to be  positive 
to make  the $x_1 = 0.5$ (i.e. $m=0$) fixed point
unstable. On the other hand, when it is stable, i.e., $2z_1 + z_2 -2  < 0$, $m =0$ is the only solution.
Fig. \ref{stability}b shows the flow diagram of the generalised voter model.

\begin{figure}[h]
\includegraphics[width=7cm]{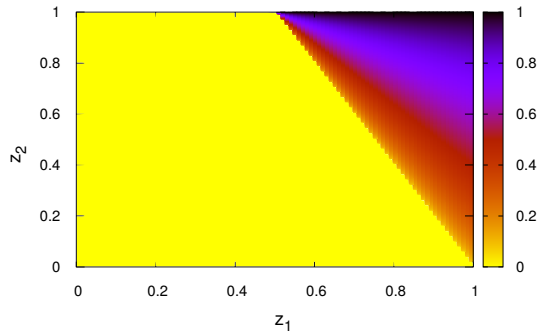}
\caption{Magnetisation $|m|$ is shown as a function of $z_1$
and $z_2$.}
\label{magnetisation}
\end{figure}

Hence the  phase boundary between the ordered and disordered phases is given by $2z_1+z_2-2 = 0$. 
We have plotted the phase diagram in Fig. \ref{magnetisation}, where the magnitude of the magnetisation is also 
indicated.  Obviously, the mean field phase diagram shows a larger region that is  ordered compared to the
two dimensional case. 

\section{Summary and Conclusions}\label{Summary}

We have studied the dynamics in zero temperature Ising-like systems  with up/down symmetry with different coordination number $z$.
Using mean field approximation,  it is observed that 
the dynamics  always lead to one unstable fixed point which corresponds to the  disordered state for $z > 2$.  
This fixed point is precisely $x=0.5$ where $x$ is the fraction of up spins. 
The stability of this fixed point has been considered by introducing a small 
deviation $\delta_0$ from  0.5 in $x$. This essentially  means we have a biased initial condition  in the 
system with unequal    fractions of   
up and down spins.  
The initial bias is generally considered to be small  such that the system does not have any appreciable correlation. 

For the unstable fixed points     
we obtain  an  initial exponential growth of $\delta(t) $ with  time which strongly depends on the coordination 
number $z$.  The growth is characterised by an exponent $\gamma$ that increases with $z$. 
These results have been checked by numerical simulations for the mean field Ising model.

The  simulations of  the short ranged Ising model in two dimensions on the other hand showed 
that  the behaviour of  $\delta(t)$ is a power law with time. The power law exponent
is non-universal and depends on $\delta_0$. 
The exit probability has also been calculated which for the two  dimensional Ising model shows the expected 
nonlinear behaviour. The exponent $\nu$ and the scaling factor $\lambda$ related to the 
finite size behaviour 
 have been  calculated. It appears that   $\nu$ shows a weak dependence on the lattice structure (i.e., $z$) while for   $\lambda$ the values are appreciably different  (Table \ref{table}). 
The exit probability study for the mean field model on the other hand shows 
$\nu^\prime$ is independent of $z$ while $\lambda$ again shows strong dependence. 
 
We have also conducted a similar study for the two parameter generalised voter model. In this case, we find that 
the stability of the disordered fixed point depends on the parameter values and it is possible to obtain a phase diagram 
based on this analysis. 

Our studies show that the behaviour of $\delta (t)$ which is related to magnetisation for
the fixed point $x= 0.5$ is different in the mean field case and the short range model. However, 
when the number of neighbours $z=2$, the mean field result that the dynamics conserve the ensemble 
magnetisation is the same as that of the  one dimensional Ising model or the voter model. 
We have justified this result on the basis of the voter model dynamics. 
Hence an important conclusion is that for $z=2$, the results are independent of the range of the interaction.

The instability of the $x= 0.5$ fixed point for the higher values of $z$ indicates the exit probability 
should be a step function in the mean field case in the thermodynamic limit.  This behaviour is found to be true  for the short range models as well in which the exit probability
for larger lattice sizes show the tendency to approach a step function.  
However, the exponents associated with the finite size scaling analysis are quite different quantitatively. 
In particular,
$\nu^\prime$  is independent of 
$z$ in the mean field case and less compared to the value obtained for the model on two dimensional lattices.

Acknowledgements: We thank B. K. Chakrabarti, P. Ray and Soham Biswas for discussions.
PS acknowledges the financial support from SERB project MTR/2020/000356 and RR thanks University of Calcutta for the University Research Fellowship (sanction no: DPO/50/Fellow(Univ)).

\end{document}